\documentclass[preprint,aps,prb,showpacs,nofootinbib,preprintnumbers,amsmath,amssymb]{revtex4-1}
\usepackage{amsfonts}
\usepackage[breaklinks=true,colorlinks,citecolor=blue,linkcolor=blue,urlcolor=blue]{hyperref}
\usepackage{subfigure,amsfonts, latexsym}
\usepackage[notcite,color,final]{showkeys}
\usepackage{epsfig,graphicx,subfigure,dcolumn,bm}
\usepackage[usenames ,dvipsnames]{xcolor}
\usepackage{slashed}
\usepackage{ulem}

\begin{document}
\title{Topological Dirac Semimetal Phase in Bismuth Based Anode Materials for Sodium-Ion Batteries}

\author{Wei-Chi Chiu}
\email{chiu.w@husky.neu.edu}
\affiliation{Department of Physics, Northeastern University, Boston, Massachusetts 02115, USA}

\author{Bahadur Singh}
\email{bahadursingh24@gmail.com}
\affiliation{Department of Physics, Northeastern University, Boston, Massachusetts 02115, USA}

\author{Sougata Mardanya}
\affiliation{Department of Physics, Indian Institute of Technology Kanpur, Kanpur 208016, India}

\author{Johannes Nokelainen}
\affiliation{Department of Physics, School of Engineering Science, Lappeenranta University of Technology, FI-53851 Lappeenranta, Finland}

\author{Amit Agarwal}
\affiliation{Department of Physics, Indian Institute of Technology Kanpur, Kanpur 208016, India}

\author{Hsin Lin}
\affiliation{Institute of Physics, Academia Sinica, Taipei 11529, Taiwan}

\author{Christopher Lane}
\affiliation{Theoretical Division, Los Alamos National Laboratory, Los Alamos, NM 87545, USA}
\affiliation{Center for Integrated Nanotechnologies, Los Alamos National Laboratory, Los Alamos, NM 87545, USA}

\author{Katariina Pussi}
\affiliation{Department of Physics, School of Engineering Science, Lappeenranta University of Technology, FI-53851 Lappeenranta, Finland}

\author{Bernardo Barbiellini}
\affiliation{Department of Physics, School of Engineering Science, Lappeenranta University of Technology, FI-53851 Lappeenranta, Finland}

\author{Arun Bansil}
\affiliation{Department of Physics, Northeastern University, Boston, Massachusetts 02115, USA}


\begin{abstract}
Bismuth has recently attracted interest in connection with Na-ion battery anodes due to its high volumetric capacity. It reacts with Na to form Na$_3$Bi which is a prototypical Dirac semimetal with a nontrivial electronic structure. Density-functional-theory based first-principles calculations are playing a key role in understanding the fascinating electronic structure of Na$_3$Bi and other topological materials. In particular, the strongly-constrained-and-appropriately-normed (SCAN) meta-generalized-gradient-approximation (meta-GGA) has shown significant improvement over the widely used generalized-gradient-approximation (GGA) scheme in capturing energetic, structural, and electronic properties of many classes of materials. Here, we discuss the electronic structure of Na$_3$Bi within the SCAN framework and show that the resulting Fermi velocities and {\it s}-band shift around the $\Gamma$ point are in better agreement with experiments than the corresponding GGA predictions. SCAN yields a purely spin-orbit-coupling (SOC) driven Dirac semimetal state in Na$_3$Bi in contrast with the earlier GGA results. Our analysis reveals the presence of a topological phase transition from the Dirac semimetal to a trivial band insulator phase in Na$_{3}$Bi$_{x}$Sb$_{1-x}$ alloys as the strength of the SOC varies with Sb content, and gives insight into the role of the SOC in modulating conduction properties of Na$_3$Bi.
\end{abstract}

\maketitle

\section{Introduction}\label{intro}

Since lithium is a nonrenewable resource~\cite{olivetti2017}, its widespread use in Li-ion batteries can be expected to lead to increasing costs of batteries in the coming years. This has motivated extensive research on Na-ion batteries~\cite{nayak2018,mukherjee2019} as an alternative to Li based batteries. However, since Na ions have a larger size and greater weight compared to Li ions, they diffuse with greater difficulty through common electrode materials. It is important therefore to develop electrode materials with a high reversible capacity and good conducting properties~\cite{luo2016}. This effort can benefit from  first-principles computations within the framework of the density functional theory (DFT)~\cite{he2019}.

The generalized gradient approximation (GGA) has been extensively used for identifying many classes of topological materials~\cite{RevModPhys.88.021004,PhysRev.136.B864, PhysRev.140.A1133, PhysRevLett.77.3865} and their novel applications~\cite{savage2018} including electrodes for Li-ion batteries~\cite{Liu2017_Topo}. The presence of symmetry-enforced Dirac states in a topological material can provide robust carriers for high electronic conductivity, which is an important factor for improved battery performance. Despite its success in predicting the first topological insulator~\cite{Zhang:2009aa}, the Dirac semimetal~\cite{PhysRevB.85.195320}, and the Weyl semimetal~\cite{Xu613}, the GGA suffers from fundamental shortcomings in describing the structural and electronic properties of materials. In~this connection, recent advances in constructing new classes of exchange-correlation functionals show that the strongly-constrained-and-appropriately-normed (SCAN)~\cite{PhysRevLett.115.036402} meta-GGA functional provides a~systematic improvement over the GGA in diversely bonded materials. SCAN is the first meta-GGA that satisfies all of the 17 exact constraints that a meta-GGA can satisfy. SCAN has been shown to yield improved modeling of metal surfaces~\cite{PatraE9188}, 2D atomically thin-films beyond graphene~\cite{Buda:2017aa}, the noncollinear antiferromagnetic ground state of manganese~\cite{pulkkinen2020}, magnetic states of copper oxide superconductors~\cite{PhysRevB.98.125140, furness2018, zhang2020}, and cathode materials for Li ion batteries ~\cite{hafiz2019}, among others materials.  

Among the various anode materials, antimony was recently shown to be a good candidate for sodium-ion batteries with strong sodium cyclability~\cite{he2018Sb, darwiche2018} and a high theoretical capacity of $660$\,mAh/g corresponding to the Na$_3$Sb phase. However, DFT calculations predict that Na$_3$Sb is prone to being an insulator~\cite{PhysRevLett.113.256403}. 
\footnote{The electronic conductivity  of the anode could be activated, for example, by using carbon matrices~\cite{Nagulapati_2019}.}
Na$_3$Bi, on the other hand, is a three-dimensional (3D) nontrivial Dirac semimetal~\cite{satya2015}, and a number of recent studies discuss Na$_3$Bi  as an anode material in sodium-ion batteries ~\cite{sottmann2016,lim2017, wang2017Bi, sun2018, huang2019}. Huang et al.~\cite{huang2019} identify a variety of different phases of  Na$_3$Bi such as NaBi, c-Na$_3$Bi (cubic), and h-Na$_3$Bi (hexagonal) as being involved in the sodiation process.

Notably, h-Na$_3$Bi (henceforth Na$_3$Bi, for simplicity) is a 3D analog of graphene. It hosts symmetry-protected, four-fold degenerate band-touching points or nodes in its bulk energy spectrum around which the energy dispersion is linear in all momentum space directions~\cite{RevModPhys.90.015001, ncomms5898, PhysRevB.93.205109, Weng_2016, PhysRevLett.108.140405}. The~aforementioned nodes or Dirac points are robust in the sense that they cannot be removed without breaking space-group symmetries. Na$_{3}$Bi is a prototypical band inversion Dirac semimetal which was predicted theoretically before its Dirac semimetal character was verified experimentally \cite{PhysRevB.85.195320,Liu864, Xu294, PhysRevB.89.245201,PhysRevB.94.085121,PhysRevB.96.075112}. The Dirac points in this material are protected by $C_3$ rotational symmetry and lie along the hexagonal $z$-axis. GGA-based studies indicate that the band inversion in Na$_3$Bi is driven essentially by crystal-field effects and does not require the presence of spin-orbit-coupling (SOC) effects, although the SOC is responsible for opening up gaps in the energy spectrum everywhere except at the Dirac points. However, the GGA has well-known shortcomings in predicting sizes of bandgaps and crystal-field splittings, providing motivation for investigating the topological structure of Na$_{3}$Bi using more advanced density functionals.       
 
Here, we revisit the topological electronic structure of Na$_3$Bi using the more accurate SCAN meta-GGA exchange-correlation functional and find that Na$_3$Bi is a trivial band insulator in the absence of the SOC in contrast to the GGA results. In our case, inclusion of the SOC drives the system into the topological Dirac semimetal state. We show that SCAN yields the band energetics and Fermi velocities in substantial agreement with the available experimental results on Na$_3$Bi. We~also discuss a~topological phase transition from the Dirac semimetal to a trivial band insulator phase in Na$_{3}$Bi$_{x}$Sb$_{1-x}$ alloys with varying Sb content. Our results establish that Na$_3$Bi is an SOC-driven topological semimetal like the common topological insulators such as Bi$_2$(Se, Te)$_3$. Since bismuth nano-sheets have been shown to display structural stability and good conduction properties after sodiation/desodiation cycles in sodium-ion batteries~\cite{huang2019}, our results further indicate the promise of Na$_3$Bi as an anode material. 

 The remainder of this paper is organized as follows. In Section~\ref{methods}, we describe computational details and discuss the crystal structure of Na$_3$Bi. The SCAN-based topological electronic properties are discussed in Section~\ref{BB_Na3Bi}. In Section~\ref{alloy}, we consider the bulk and surface electronic properties of Na$_{3}$Bi$_{x}$Sb$_{1-x}$ alloy. Finally, we present brief concluding  remarks in Section~\ref{conclusion}.
 
\section{Methodology and crystal structure} \label{methods}

Electronic structure calculations were carried out within the DFT framework with the projector augmented wave (PAW) method using the Vienna ab initio  Simulation Package (VASP)~\cite{PhysRevB.59.1758,PhysRev.136.B864,PhysRev.140.A1133,PhysRevB.54.11169,PhysRevLett.77.3865}. We used the GGA and SCAN meta-GGA energy functionals with the Perdew--Burke--Ernzerhof (PBE) parametrization~\cite{PhysRevLett.115.036402} to include exchange-correlation effects in computations. An energy cut-off of 400\,eV was used for the plane-wave-basis set and a $\Gamma$-centered $17\times17\times10$ $k$-mesh was employed to sample the bulk Brillouin zone (BZ). SOC effects were included self-consistently. The topological analysis was performed by employing a real-space tight-binding model Hamiltonian, which was obtained by using the VASP2WANNIER90 interface~\cite{PhysRevB.56.12847}. Bi \textit{p} and Na \textit{s} and \textit{p} states were included in generating Wannier functions. The surface electronic structure was calculated using the iterative Green’s function method as implemented in the WannierTools package ~\cite{WU2018405}. 

\begin{figure}[h!]
\includegraphics[width=0.6\textwidth]{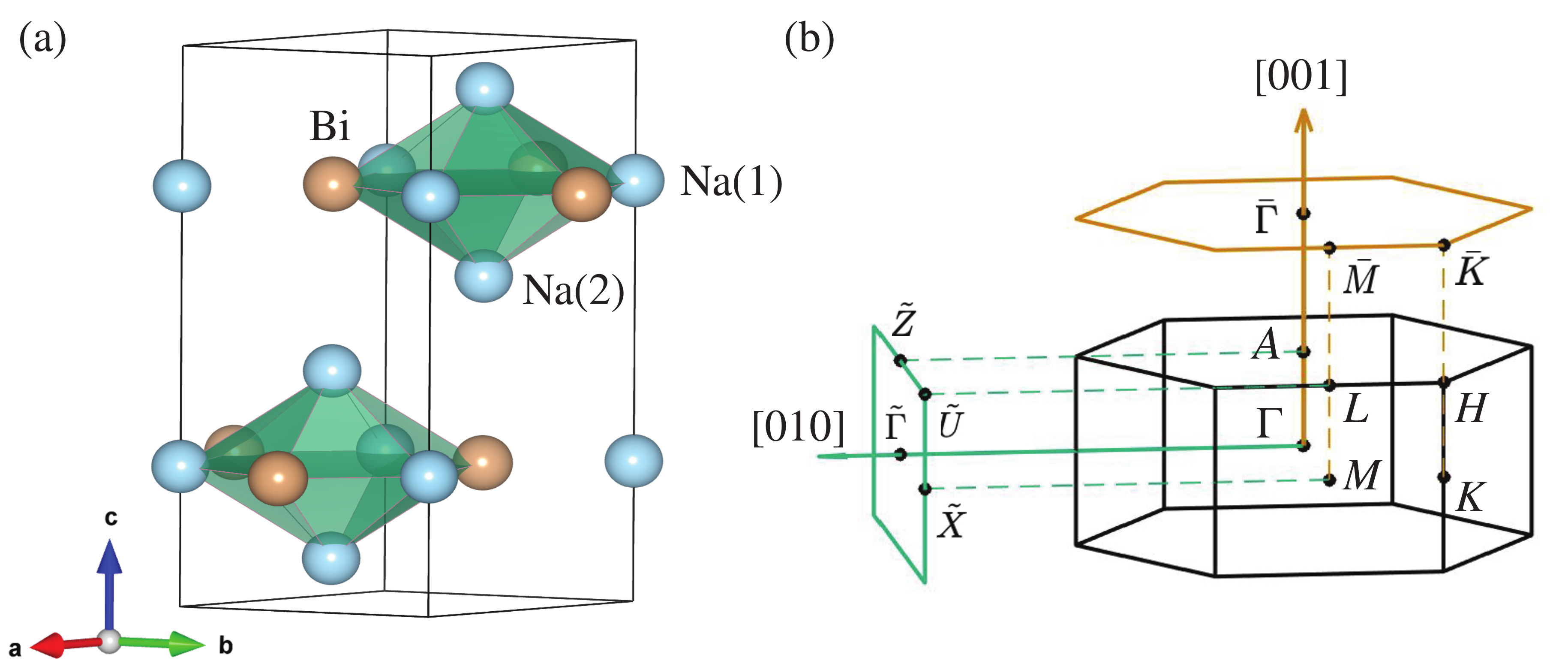}
\caption{ (\textbf{a}) crystal structure of Na$_{3}$Bi {visualized using the VESTA~\cite{Momma:db5098} package}. Na and Bi atoms are shown as blue and orange spheres. The two nonequivalent Na atoms are marked as Na(1) and Na(2); 
(\textbf{b}) bulk and  $\left[ 001 \right]$ (orange) and $\left[ 010 \right]$ (green) surface projected Brillouin zones. The relevant high-symmetry points are marked.}
\label{fig:CS}
\end{figure}

Figure~\ref{fig:CS}a shows the hexagonal crystal structure of Na$_{3}$Bi with lattice parameters  $a=5.448\,${\AA} and $c=9.655\,${\AA} and space group $D^{4}_{6h}$ ($P6_3/mmc$, No. 194).
It has a layered crystal structure where Na(1) and Bi atoms in the Wyckoff positions $2b~ [\pm(0 , 0, \frac{1}{4})]$ and $2c~ [\pm(\frac{1}{3} , \frac{2}{3}, \frac{1}{4})]$ form a~shared honeycomb structure.  The Na(2) atoms with Wyckoff position $4f ~[\pm(\frac{1}{3}, \frac{2}{3}, u)$  and $\pm(\frac{2}{3}, \frac{1}{3}, \frac{1}{2}+u);u=0.583]$ form a triangular lattice which is inserted between the honeycomb layers along the $z$-axis. Here, Na(1) and Na(2) represent two nonequivalent Na atoms in the unit cell. The bulk and surface BZs are shown in~Figure~\ref{fig:CS}b.


\section{SOC-Driven Topological Dirac Semimetal}\label{BB_Na3Bi}

\begin{figure}[t]
\includegraphics[width=0.75\textwidth]{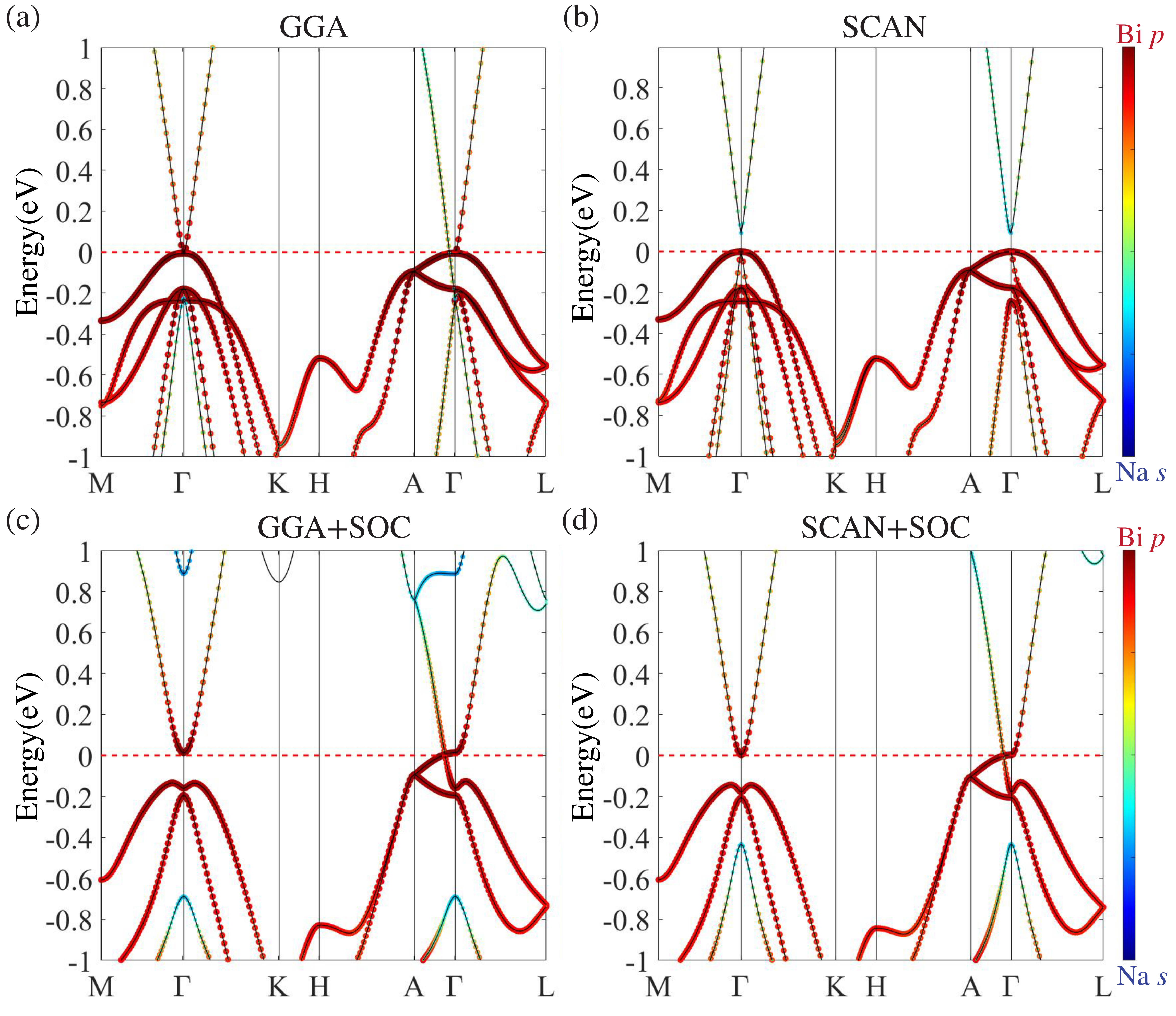}
\caption{ (Bulk band structure of Na$_{3}$Bi obtained (without spin-orbit-coupling (SOC)) using (\textbf{a}) GGA 
and (\textbf{b}) SCAN meta-GGA. The Na $s$ and Bi $p$ states are shown as blue and red markers; (\textbf{c}) and (\textbf{d}) are same as (a) and (b) except that the SOC is included in the computations.}
\label{fig:bulkbands}
\end{figure}

The bulk electronic structure of Na$_3$Bi without SOC obtained with GGA is shown in Figure~\ref{fig:bulkbands}a. It~unveils a band inversion semimetal state in which the doubly-degenerate Bi $p_{xy}$ and singly-degenerate Bi $p_z$ states cross along the $\Gamma-A$ direction and form two triply-degenerate points. At $\Gamma$, Na $s$ states are inverted and located $0.3$\,eV below the Bi $p_{xy}$ states. Figure~\ref{fig:bulkbands}c presents the band structure with SOC. The nodal states now become gapped everywhere except at two discrete points along the $\Gamma-A$ line ($C_{3}$ rotational axis). The strength of band inversion is enhanced such that Na $s$ states are lowered to $0.7$ eV below the Bi $p_{xy}$ states. 

\begin{figure}[h!]
\includegraphics[width=0.89\textwidth]{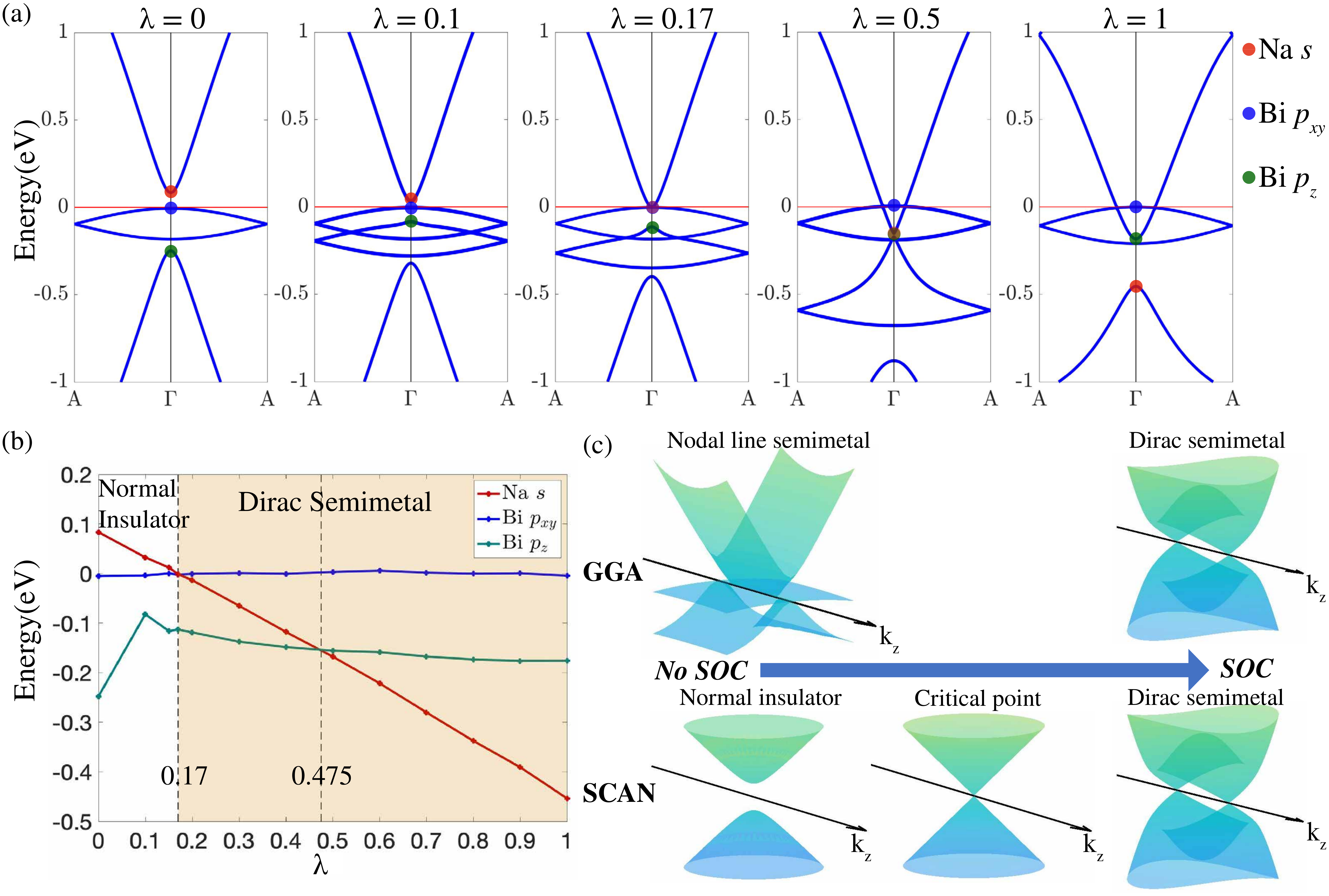}
\caption{ (\textbf{a}) energy bands in Na$_3$Bi along the $A-\Gamma-A$ symmetry line as the strength $\lambda$ of the SOC is varied from 0 to 1. Red, blue, and green dots mark the Na-$s$, Bi-$p_{xy}$, and Bi-$p_z$ derived levels, respectively, at $\Gamma$; (\textbf{b}) energies of the Na $s$ (red), Bi $p_{xy}$ (blue), and Bi $p_z$ (green) levels at $\Gamma$ as a function of $\lambda$. Orange shading marks the Dirac semimetal region; (\textbf{c})  a schematic of how the Dirac semimetal forms in GGA (top) and SCAN (bottom) as $\lambda$ is varied. } 
\label{fig:socvary}
\end{figure}

Figure~\ref{fig:bulkbands}b shows the energy bands without SOC calculated using SCAN. An insulating ground state with a band gap of 90 meV is seen in sharp contrast to the GGA results. When the SOC is included, SCAN yields an inverted band structure and the Dirac semimetal state with a pair of Dirac points (Figure~\ref{fig:bulkbands}d), although the band inversion strength is reduced and the Na $s$ states are lifted to lie 0.45 eV below the Bi $p_{xy}$ states. The shift of the Na $s$ band can be verified via angle-resolved photoemission spectroscopy (ARPES) experiments. Since the band inversion in Na$_3$Bi is driven by SOC, our SCAN-based results suggest that it should be possible to realize a topological phase transition from normal insulator to Dirac semimetal by modulating the strength of the SOC. In this connection, Figure~\ref{fig:socvary}c illustrates how the GGA and SCAN based band structures can be expected to evolve in Na$_3$Bi as a function of the strength of the SOC. 

We turn now to discuss details of the Dirac points and the related Fermi velocities and compare our theoretical predictions with the corresponding experimental results. We obtained the Fermi velocity $\nu = (\nu_x, \nu_y, \nu_z)$, where $\nu_x$, $\nu_y$, and $\nu_z$ are the velocities along the $k_x$, $k_y$, and $k_z$ directions in the bottom half of the Dirac cone, respectively, via a linear fit to the band structures over the momentum range of $-0.2$\,\AA$^{-1}$ to 0.2\,\AA$^{-1}$. The resulting values are: $\nu = (2.3187, 2.0645, 0.3144)\,$eV$\cdot$\r{A} for GGA and $\nu=(2.6196, 2.3039, 0.3199)$\,eV$\cdot$\r{A} for SCAN. The Fermi velocity obtained from ARPES measurements~\cite{Liu864} is $\nu=(2.75, 2.39, 0.6)$\,eV$\cdot$\r{A}.\footnote{(Error bars on the experimental velocities are not reported in Ref.~\cite{Liu864}.} Locations of the Dirac points $\mathbf{k}_\mathrm{d}$ are $(0, 0, \pm 0.26\,\frac{\pi}{c})$,  $(0, 0, \pm 0.2\,\frac{\pi}{c})$, from GGA and SCAN, respectively, while the corresponding experimental value is $(0, 0, \pm 0.29\,\frac{\pi}{c})$ with $\delta k_z=\pm0.17\frac{\pi}{c}$~\cite{Liu864}. The SCAN-based Fermi velocity is seen to be in better agreement with experiment. The~locations of the Dirac points from SCAN and GGA are both within the experimental resolution.

As we already pointed out, a topological phase transition in Na$_3$Bi could be realized by tuning SOC. We demonstrate this by adding a scaling factor $\lambda$ to the SOC term in the Hamiltonian as $H_\mathrm{soc} = \lambda \left(\frac{{\mu}_\mathrm{B}}{\hbar e{m }_{e}{c}^{2}}\frac{1}{r}\frac{\partial U}{\partial r} \mathbf{L}\cdot \mathbf{S} \right)$. The resulting energy dispersion obtained self-consistently along the $A -\Gamma-A$ direction is presented in Figure~\ref{fig:socvary}a for a series of $\lambda$ values. An insulator state is realized at $\lambda = 0$.  At $\lambda =0.17$, the bandgap vanishes as the Na $s$-derived conduction band crosses the Bi $p_{xy}$-derived valence band at the $\Gamma$ point, and the system reaches a topological critical point. As $\lambda$ is increased further, two Dirac points start to emerge along the $A-\Gamma-A$ line at the Fermi level. At~$\lambda=0.5$, the Na-$s$ and Bi-$p_{z}$ bands cross in the valence band continuum region to form a second critical point. Finally, at $\lambda =1$, the Na-$s$ band shifts down to 0.45\,eV below the Fermi level, while the system preserves a pair of Dirac states at the Fermi level.  Evolution of energies of the Na-$s$ and Bi-$p$ states at $\Gamma$  as a function of $\lambda$ is shown in Figure~\ref{fig:socvary}b. The Na-$s$ band evolves linearly as $\lambda$ increases from 0 to 1, but the Bi $p_{xy}$ states are seen to be almost independent of $\lambda$. In this way, the SOC realizes the band inversion by breaking the degeneracy of the doubly-degenerate Bi $p_{xy}$ states and shifting up the Bi $p_{z}$ state.

In Figure~\ref{fig:socvary}c, we illustrate schematically how the bulk electronic structure of Na$_3$Bi evolves within GGA and SCAN as the strength $\lambda$ of the SOC is varied from 0 to 1. GGA yields a nodal-line semimetal at $\lambda=0$, which evolves into a Dirac semimetal with increasing $\lambda$, so that the SOC is a~secondary effect that breaks the degeneracy of Bi $p_{xy}$ and shifts the Bi $p_{z}$ level up to invert with Na $s$ level. However, for the bands which form the Dirac points, the topology is dominated in the GGA by the crystal field which inverts the Bi $p_{xy}$ and the Na $s$ levels. If the symmetry is preserved, a topological phase transition in GGA can therefore only be achieved through an additional controlling parameter (other than the SOC) such as lattice strain along the c-axis~\cite{Guan:2017aa}. In contrast, the SOC provides sufficient control within SCAN to realize a topological phase transition. 

\section{Topological Properties of Na$_3$Bi$_x$Sb$_{1-x}$} \label{alloy}

Modulation of the SOC strength could be realized experimentally in Na$_3$Bi by forming Na$_3$Bi$_{x}$Sb$_{1-x}$ solid solutions where the SOC will weaken as the Bi atoms are replaced by the lighter Sb atoms. Along this line, we consider the end-compound Na$_3$Sb in the Na$_3$Bi structure, and find Na$_3$Sb to be a trivial insulator with SCAN-based optimized lattice parameters to be: $a=5.355$\,\r{A} and $c=9.496$\,\r{A}. Na$_3$Sb hosts a bandgap of 0.74\,eV and the electronic states around the Fermi level are derived from Na-$s$ and Sb-$p$ orbitals. Notably, SCAN gives a bandgap, which is larger than the GGA value of $\sim 0.5$\,eV~\cite{PhysRevLett.113.256403}. The SCAN bandgap is in better agreement with the experimental value of $\sim1.1$\,eV observed through absorption and photoconductivity measurements~\cite{PhysRev.112.114}.

\begin{figure}[h!]
\includegraphics[width=0.89\textwidth]{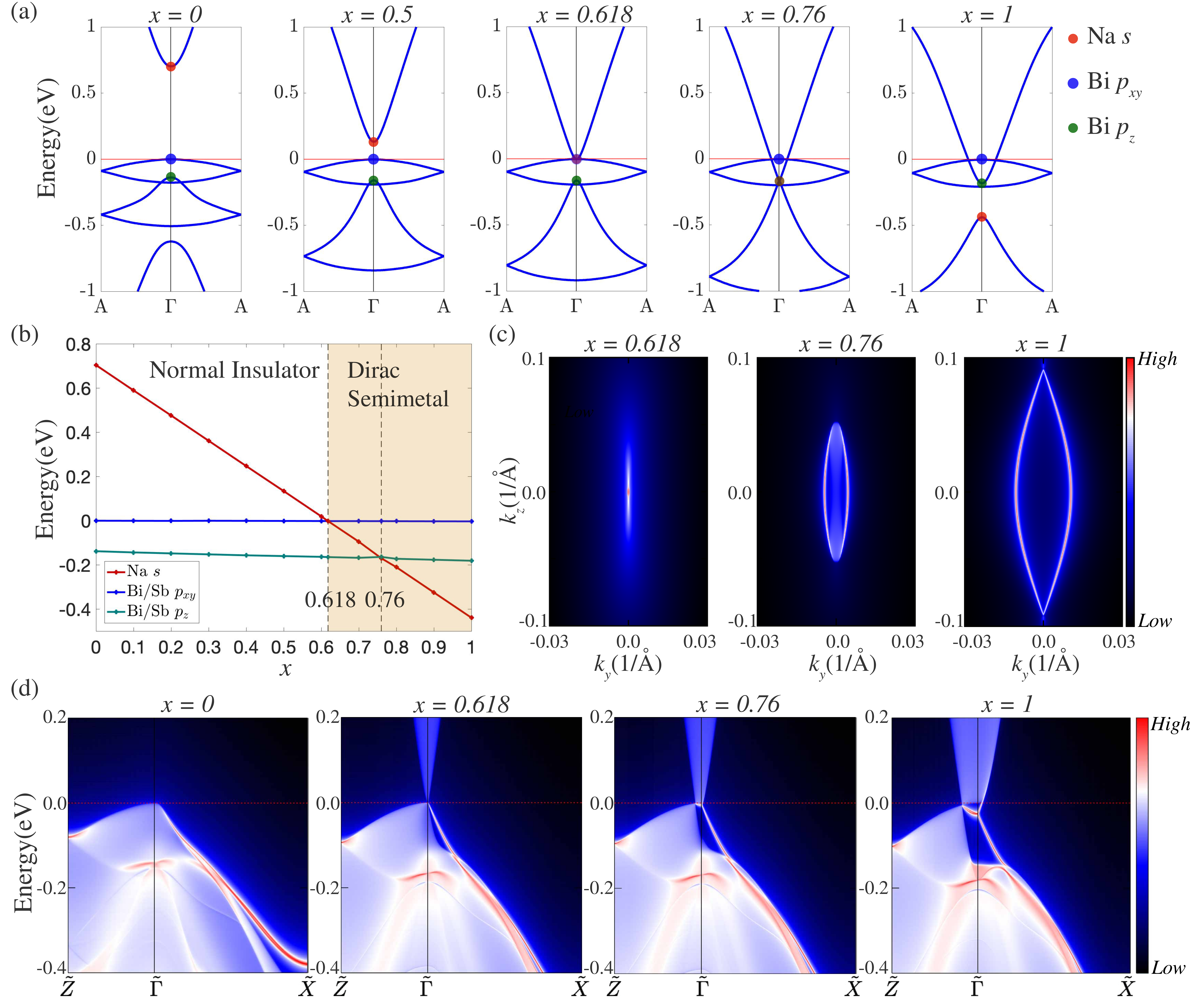}
\caption{ Bulk band structure of Na$_3$Bi$_x$Sb$_{1-x}$ alloys along the $A-\Gamma-A$ symmetry line in the BZ for various $x$ values. Red, blue, and green markers identify Na-$s$, Bi/Sb-$p_{xy}$, and Bi/Sb-$p_z$ derived levels, respectively; (\textbf{b}) evolution of the Na- and Bi/Sb-derived levels at $\Gamma$ as a function of  $x$. Shaded region marks the Dirac semimetal phase; (\textbf{c}) topological double-Fermi-arcs at $x=0.618$ (left), $x=0.76$ (middle), and $x=1$(right); (\textbf{d}) surface band structure for the [010] surface of Na$_3$Bi$_x$Sb$_{1-x}$ alloys at $x=0$, $x=0.618$, $x=0.76$, and $x=1$. } 
\label{fig:vca}
\end{figure}

We have investigated the electronic structure of  Na$_3$Bi$_{x}$Sb$_{1-x}$ alloys within the virtual-crystal-approximation (VCA), which is a reasonable description for alloys in which the dopant and host atoms have similar chemical compositions~\cite{PhysRevB.20.4025,PhysRevB.20.4035,PhysRevB.88.224202,PhysRevLett.120.136403}.
Figure~\ref{fig:vca}a shows the bulk band structure of Na$_3$Bi$_{x}$Sb$_{1-x}$ alloys for various values of $x$. The band structure in the vicinity of the Fermi level is seen to evolve with $x$ along the lines discussed above in connection with the evolution of the band structure in Na$_3$Bi with varying SOC strength. At $x=0$, there is a clear band gap between the Na-$s$ and Bi/Sb-$p_{xy}$ levels. At $x =0.618$, the gap between the valence and conduction states vanishes and the system reaches a topological critical point.  With increasing $x$, a Dirac semimetal state with a pair of Dirac points on the $A-\Gamma-A$ symmetry line emerges and the Na-$s$ derived levels move down to the Fermi level. The energy variation of various levels at $\Gamma$ as a function of $x$ is illustrated in Figure~\ref{fig:vca}b.

The SCAN-based surface band structure for the [010] surface for various values of $x$ is presented in Figure~\ref{fig:vca}d. There is no surface state connecting the valence and conduction bands at $x=0$ as expected from the bulk band structure of Na$_3$Sb. At $x=0.618$, the system reaches a critical point. With increasing $x$, the critical point splits into two bulk Dirac cones along the $A-\Gamma-A$ symmetry line, and the nontrivial surface states connecting the Dirac cones emerge and can be seen for $0.618<x<1$ as depicted in Figure~\ref{fig:vca}c. The preceding surface state results are consistent with the evolution of the bulk band topology with $x$. Although we have discussed the electronic structure of the Na$_3$Bi$_{x}$Sb$_{1-x}$ alloys using the relatively simpler VCA scheme, a more sophisticated treatment of disorder effects using the coherent-potential-approximation (CPA) will be interesting \cite{PhysRevLett.113.256403,PhysRevB.60.13396}.

\section{Conclusion}\label{conclusion}
We discuss the topological electronic structure of Na$_3$Bi using the recently developed SCAN meta-GGA functional within the first-principles  DFT framework. Our SCAN-based band structure and Fermi velocities of the Dirac states are in better accordance with the corresponding experimental results compared to the earlier GGA-based results. Nature of the Dirac states in Na$_3$Bi is examined in depth by exploring effects of the strength of the SOC on the topology of the band structure. The SCAN-based Dirac semimetal state in Na$_3$Bi is shown to be driven by SOC effects in contrast to the GGA where the topological phase appears even in the absence of the SOC. We also consider Na$_3$Bi$_x$Sb$_{1-x}$ alloys and show that a topological phase transition can be realized from the Dirac semimetal state to a trivial band insulator by varying the Bi/Sb concentration. Our analysis indicates that the topological state of pristine Na$_3$Bi is very robust compared to other anode materials such as Na$_3$Sb~\cite{PhysRevLett.113.256403} and MoS$_2$~\cite{lane2019}, which can undergo metal-insulator transitions. The presence of Dirac bands with high carrier velocities and the associated conductivity provide a favorable factor for battery performance.

\section*{ACKNOWLEDGMENTS}
The work at Northeastern University was supported by the U.S. Department of Energy (DOE), Office of Science, Basic Energy Sciences Grant No. DE-FG02-07ER46352, and benefited from Northeastern University’s Advanced Scientific Computation Center and the National Energy Research Scientific Computing Center through DOE Grant No. DE-AC02-05CH11231. J.N. is supported by the Finnish Cultural Foundation. H. L. acknowledges Academia Sinica (Taiwan) for the support under Innovative Materials and Analysis Technology Exploration (AS-iMATE-107-11).

\bibliography{Na3Bi}

\end{document}